\documentclass[twocolumn,showpacs,preprintnumbers,amsmath,amssymb]{revtex4}

\usepackage{graphicx}
\usepackage{graphicx}
\usepackage{dcolumn}
\usepackage{bm}

\begin{document}

\title{Enhance synchronizability via age-based coupling}

\author{Yu-Feng Lu$^1$}
\author{Ming Zhao$^1$}
\email{zhaom17@mail.ustc.edu.cn}
\author{Tao Zhou$^{1,2}$}
\email{zhutou@ustc.edu}
\author{Bing-Hong Wang$^1$}
\email{bhwang@ustc.edu.cn} \affiliation{$^1$ Department of Modern
Physics, University of Science and Technology of China, Hefei
230026, China\\ $^2$ Department of Physics, University of Fribourg,
CH-1700 Fribourg, Switzerland}
\date{\today}

\begin{abstract}
In this brief report, we study the synchronization of growing
scale-free networks. An asymmetrical age-based coupling method is
proposed with only one free parameter $\alpha$. Although the
coupling matrix is asymmetric, our coupling method could guarantee
that all the eigenvalues are non-negative reals. The eigneratio $R$
will approach to 1 in the large limit of $\alpha$.
\end{abstract}


\maketitle

One of the main goals in studying of network science is to
understand the relation between network structure and dynamical
processes performed upon \cite{Newman3,Boccaletti}. A typical
collective dynamic on networked system is synchronization, where all
the participants behave alike, even exactly the same. This
phenomenon exists everywhere from physics to biology \cite{E. M},
and has been observed for hundreds of years. With the partial
knowledge of relations between network structure and its
synchronizability \cite{Nishikawa03,Hong04,ZhaoM2006,McGraw05,Wu06},
scientists have proposed many methods to enhance the network
synchronizability
\cite{ZhaoM05,ZhouT06CDC,YinCY06,Motter05,MZKEPL,MZKAIP,Hwang05,Chavez05,ZhouC06,ZhaoM06EPJB}.
Generally speaking, these methods can be divided into two classes,
one is to modify the network structure
\cite{ZhaoM05,ZhouT06CDC,YinCY06}, the other is to regulate the
coupling pattern
\cite{Motter05,MZKEPL,MZKAIP,Hwang05,Chavez05,ZhouC06,ZhaoM06EPJB}.
In the former class, networks are modified either to shorten the
average distance \cite{ZhouT06CDC} or to eliminate the maximal
betweenness \cite{ZhaoM05,YinCY06}. In the later case, the network
structure is kept unchanged, while the coupling matrix is
elaborately designed (often asymmetric) to improve the
synchronizability
\cite{Motter05,MZKEPL,MZKAIP,Hwang05,Chavez05,ZhouC06,ZhaoM06EPJB}.

The first coupling pattern other than the symmetric case was
proposed by Motter-Zhou-Kurths \cite{Motter05,MZKEPL,MZKAIP} (MZK
coupling pattern), in which the coupling strength a node $i$
receives from its neighbors is inverse to $k^{\beta}_i$ with $k_i$
the degree of $i$. The coupling pattern can sharply enhance the
network synchronizability, with $\beta=1$ the optimal case. After
this pioneer work, many coupling patterns
\cite{Hwang05,Chavez05,ZhouC06,ZhaoM06EPJB} have been presented to
further enhance the network synchronizability. In Ref.
\cite{Hwang05}, Hwang \emph{et al.} presented a coupling method
taking into account the age of nodes, which makes the network even
more synchronizable than the optimal case of MZK coupling pattern.
In this pattern, each node receives coupling signals from its
neighbors, with each receiving coupling strength taking one of the
two values: if the neighbor is older, the coupling strength takes
the larger value, otherwise it takes the smaller one. To separate
the different coupling situations (i.e. from older to younger and
from younger to older) by using two different coupling strengths is
the simplest way one can image. However, since each node has its own
age, a coupling method taking into account the age difference
between each pair of coupled nodes may further enhance the
synchronizability. Moreover, the coupling matrix in Ref.
\cite{Hwang05} has complex eigenvalues, leading to a complicated
analysis. An elaborately designed method, as shown in this brief
report, could guarantee that all the eigenvalues are nonnegative
reals, thus one can easily predict the synchronizability of
underlying network by considering the real eigneratio only.

In a dynamical network, each node represents an oscillator and the
edges represent the couplings between nodes. For a network of $N$
linearly coupled identical oscillators, the dynamical equation of
each oscillator can be written as
\begin{equation}
\dot{\textbf{x}}^i=\textbf{F}(\textbf{x}^i)-\sigma\sum_{j=1}^NG_{ij}\textbf{H}(\textbf{x}^j),\hspace*{1em}i=1,2,...,N,
\end{equation}
where $\dot{\textbf{x}}^i=\textbf{F}(\textbf{x}^i)$ governs the
essential dynamics of the $i$th oscillator,
$\textbf{H}(\textbf{x}^j)$ the output function, $\sigma$ the
coupling strength, and $G_{ij}$ an element of the $N\times N$
coupling matrix $G$. To guarantee the synchronization manifold an
invariant manifold, the matrix $G$ should has zero row-sum. The
collective dynamic starts from a disorder initial configuration,
under suitable conditions, the couplings will make the oscillators'
states nearer and nearer. Eventually, all the individuals oscillate
together, and synchronization phenomenon emerges.

In the simplest symmetric way, the coupling matrix $G$ has the same
form as the Laplacian matrix $L$, that is, $G_{ij}=L_{ij}$ where
\begin{equation}
    L_{ij}=\left\{
    \begin{array}{cc}
    k_i   &\mbox{for $i=j$}\\
     -1    &\mbox{for $j\in\Lambda_i$}   \\
     0    &\mbox{otherwise}
    \end{array}
    \right..
\end{equation}
Here $\Lambda_i$ is the set of $i$'s neighbors. Because of the
symmetry and the positive semidefinite of $L$, all its eigenvalues
are nonnegative reals and the smallest eigenvalue $\lambda_0$ is
always zero, for the rows of $L$ have zero sum. And if the network
is connected, there is only one zero eigenvalue. Thus, the
eigenvalues can be ranked as
$\lambda_0<\lambda_1\leq\lambda_2\leq...\leq\lambda_{N-1}$. When the
stability zone is bounded, according to the criteria of master
stability function \cite{Pecora98,Pecora02} (see also the unbounded
case \cite{Wang2002a,Wang2002b}), the network synchronizability can
be measured by the eigenratio $R=\lambda_{N-1}/\lambda_1$: The
smaller it is the better the network synchronizability and vice
versa.

The couplings between nodes are not limited to the symmetric mode,
however, generally, the eigenratio of an asymmetric coupling matrix
is complex (e.g. the eigenratio in Ref. \cite{Hwang05}). Therefore,
in order to ensure the network having strong synchronizability, not
only the ratio of the real part should be taken into account, but
also the imaginary part must be guaranteed as small as possible. In
Ref. \cite{Hwang05}, the simulation result indicated that although
the ratio of the real part is the smallest, at the same time the
imaginary part is the largest. To overcome this blemish and give
further enhancement of synchronizability, we bring forward a
coupling pattern in which the coupling strength between two
connected nodes is the function of their age difference. The age of
node $i$ is signed by the time it enters into the network, thus
smaller $i$ corresponds to older age. The coupling matrix proposed
here is:
\begin{equation}
    G_{ij}=\left\{
    \begin{array}{cc}
    1   &\mbox{for $i=j$}\\
     -\frac{e^{-\frac{\alpha}{N}(j-i)}}{S_i}    &\mbox{for $j\in\Lambda_i$}   \\
     0    &\mbox{otherwise}
    \end{array}
    \right.,
\end{equation}
where $S_i=\sum_{j\in \Lambda_i}e^{-\frac{\alpha}{N}(j-i)}$ is the
normalization factor. In this coupling pattern, the case of
$\alpha=0$ degenerates to the optimal case of MZK coupling pattern.
When $\alpha>0$, the old nodes have stronger influence than the
younger ones; while for $\alpha<0$, younger nodes are more
influential.

It can be proved that although the coupling between nodes is
asymmetric, all the eigenvalues of matrix $G$ are reals. Note that,
the coupling matrix defined in (3) can be written as
\begin{equation}
G=DL',
\end{equation}
where
\begin{equation}
D=\mbox{diag}(e^{2\alpha}/S_{1}, e^{4\alpha}/S_{2},
e^{6\alpha}/S_{3},..., e^{2N\alpha}/S_{N})
\end{equation}
is a diagonal matrix, and $L'=(L_{ij}')$ is a symmetric zero row-sum
matrix, whose off-diagonal elements are
\begin{equation}
L'_{ij}=-e^{-\alpha i}e^{-\alpha j}.
\end{equation}
From the identity
\begin{equation}
\mbox{det}(DL'-\lambda
I)=\mbox{det}(D^{\frac{1}{2}}L'D^{\frac{1}{2}}-\lambda I)
\end{equation}
valid for any $\lambda$, we have that the spectrum of eigenvalues of
matrix $G$ is equal to the spectrum of a symmetric matrix defined as
\begin{equation}
H=D^{\frac{1}{2}}L'D^{\frac{1}{2}}.
\end{equation}
As a result, although the coupling matrix $G$ is asymmetry, the
eigenvalues of matrix $G$ are all nonnegative reals and the smallest
eigenvalue is always zero. Therefore, different from the complicated
case in Ref. \cite{Hwang05}, the synchronizability based on the
present coupling pattern can be measured directly by the real
eigenratio $R$.

\begin{figure}
\scalebox{0.85}[0.85]{\includegraphics{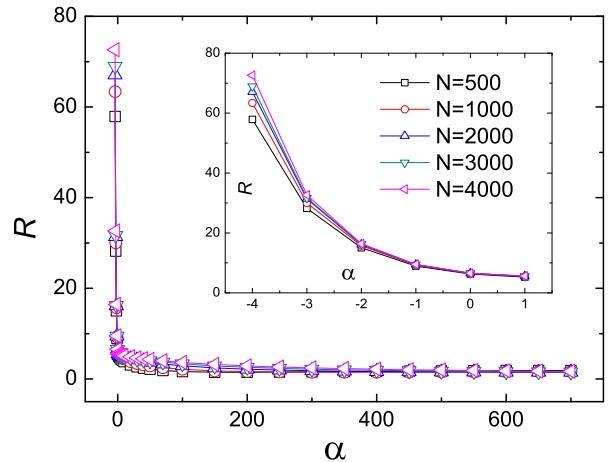}}
\caption{\label{fig:epsart} (Color online) The eigenratio $R$ vs.
$\alpha$ in BA networks with average degree $\langle k\rangle=6$.
The inset displays the details for the interval $\alpha \in [-4,1]$.
Each data point is obtained by averaging over 50 different network
configurations. The eigenratio $R$ goes to 1 in the large limit of
$\alpha$. }
\end{figure}

In Fig. 1, we report the changes of eigenratio $R$ with the
parameter $\alpha$ in BA networks \cite{Brabasi2} at different
sizes. One can easily conclude from Fig. 1 that with the increasing
of $\alpha$ the eigenration decreases sharply, no matter what the
network size is. It is shown that in growing networks, if the
couplings from older nodes are stronger than the reverse, the
network will get better synchronizability. Otherwise, if the
coupling from younger to older ones is strengthened (see the cases
of $\alpha<0$ in the inset), the system becomes very hard to
synchronize. When $\alpha$ goes to infinite, the eigenratio will
converge to 1, which is the possibly smallest eigenratio
corresponding to the best synchronizability \cite{Nishikawa06}.
Actually, in the case $\alpha\rightarrow+\infty$, each node is
coupled by its oldest neighbor, while the oldest node in the network
is uncoupled. Thus, the coupling matrix (whose rows are sorted by
the descending order of ages) becomes a lower triangular matrix with
all the diagonal elements are 1 except the first one, $G_{11}$,
being equal to zero. Therefore, all the non-zero eigenvalues are
one.

Although there exists various methods to design a coupling pattern
having optimal synchronizability (i.e. $R=1$) \cite{Nishikawa06},
for growing networks, using the age of each node is a simple and
feasible way since to know any other measures of nodes may cost much
for huge-size system, and this age-based coupling can guarantee the
connectivity of the whole network. Mathematically speaking, the
synchronizability here is a measure on the stability of invariant
synchronization manifold. We call a synchronization manifold is
stable if the dynamical system can automatically return to this
manifold after a perturbation. A network $G$ has better
synchronizability than another network $G'$ means any collective
dynamics with identical oscillators upon $G'$ having a stable
synchronization manifold must have a stable synchronization manifold
for $G$, while there exists certain dynamics having stable
synchronization manifold for $G$, but not for $G'$. However, better
synchronizability does not guarantee a shorter converging time from
disorder initial configuration to synchronized state. Actually,
Nishikawa and Motter \cite{Nishikawa06} found that the synchronizing
process may take longer time in the optimal network with $R=1$ (see
also a similar conclusion for non-identical oscillators
\cite{Um2007}). Based on the current coupling pattern, one can
obtain his acceptable trade-off between synchronizability and
converging time by tuning the parameter $\alpha$. Moreover,
comparing with the pioneer work by Hwang \emph{et al.}
\cite{Hwang05}, our coupling method can achieve even smaller $R$,
and does not need to deal with the complicated and boring analysis
on complex eigenratio. Instead, our elaborately designed coupling
pattern can guarantee the eigenratio a real number.

This work is partially supported by NNSFC under Grant Nos. 10472116,
10635040 and A0524701, as well as the specialized program under
President Funding of Chinese Academy of Science.

\end{document}